\documentclass[
superscriptaddress,
 amsmath,
 amssymb,
 aps,
 pra,
floatfix,
12pt]{revtex4-2}

\usepackage{graphicx}
\usepackage{dcolumn}
\usepackage{bm}
\usepackage{physics}
\usepackage{natbib}
\usepackage{color}

\begin{document}


\title{Quantum undetected optical projection tomography}

\author{Nathan~R.~Gemmell}
\email{n.gemmell20@imperial.ac.uk}
\author{Emma~Pearce}%
\author{Jefferson~Fl\'orez}%

\author{Rupert~F.~Oulton}%
\affiliation{%
 Department of Physics, Blackett Laboratory, Imperial College London, South Kensington Campus, London SW7 2AZ, United Kingdom
}%

\author{Alex~S.~Clark}
\affiliation{%
 Department of Physics, Blackett Laboratory, Imperial College London, South Kensington Campus, London SW7 2AZ, United Kingdom
}%
\affiliation{
 Quantum Engineering Technology Labs, H. H. Wills Physics Laboratory and School of Electrical, Electronic and Mechanical Engineering, University of Bristol, BS8 1FD, United Kingdom
}%
\author{Chris~C.~Phillips}%
\affiliation{%
 Department of Physics, Blackett Laboratory, Imperial College London, South Kensington Campus, London SW7 2AZ, United Kingdom
}%

\date{\today}

\begin{abstract}

Quantum imaging with undetected photons (QIUP) is an emerging technique that decouples the processes of illuminating an object and projecting its image. The properties of the illuminating and detected light can thus be simultaneously optimised for both contrast on a sample and sensitivity on a camera. Here, we combine QIUP with computed tomography to enable three-dimensional (3D) infrared imaging. The image data is registered with a standard silicon camera at a wavelength of 810\,nm, but the extracted 3D images map the sample's absorption at a wavelength of 1550\,nm, well beyond the camera’s sensitivity. Quantum Undetected Optical Projection Tomography (QUOPT) enables label-free volumetric sensing at difficult to detect wavelengths, such as those that allow molecular imaging contrast, or those within the infrared biological transmission windows.

\end{abstract}

\maketitle


\section{\label{sec:level1}Introduction}

Computed tomography (CT) \cite{hounsfield1973computerized,scudder1978introduction} is a powerful three-dimensional (3D) mapping procedure, commonly recognised through its use in x-ray medical imaging \cite{bull1980history,rubin2014computed,shepp1978computerized}. The transmissivity of a subject is measured at a range of known angles, and a computerised reconstruction then allows the transmission losses to be mapped within the scanned 3D volume. In medicine and healthcare it allows knowledge of a patient's internal structure to be obtained non-invasively \cite{rawson2020x, garvey2002computed}, and it has also found a number of other non-medical uses such as determining the internal structures of materials \cite{withers2021x,van2005computed}, geological/paleontological samples \cite{cnudde2013high, carlson2003applications}, and plant structures \cite{stuppy2003three}. The CT approach has also been extended to the visible part of the electromagnetic spectrum, termed optical projection tomography (OPT) \cite{sharpe2004optical}, where it has been used, for example, to image embryonic organs \cite{colas2009live} and live gene activity in plants \cite{lee2017macro}.

\begin{figure*}[t]
\includegraphics[width=\textwidth]{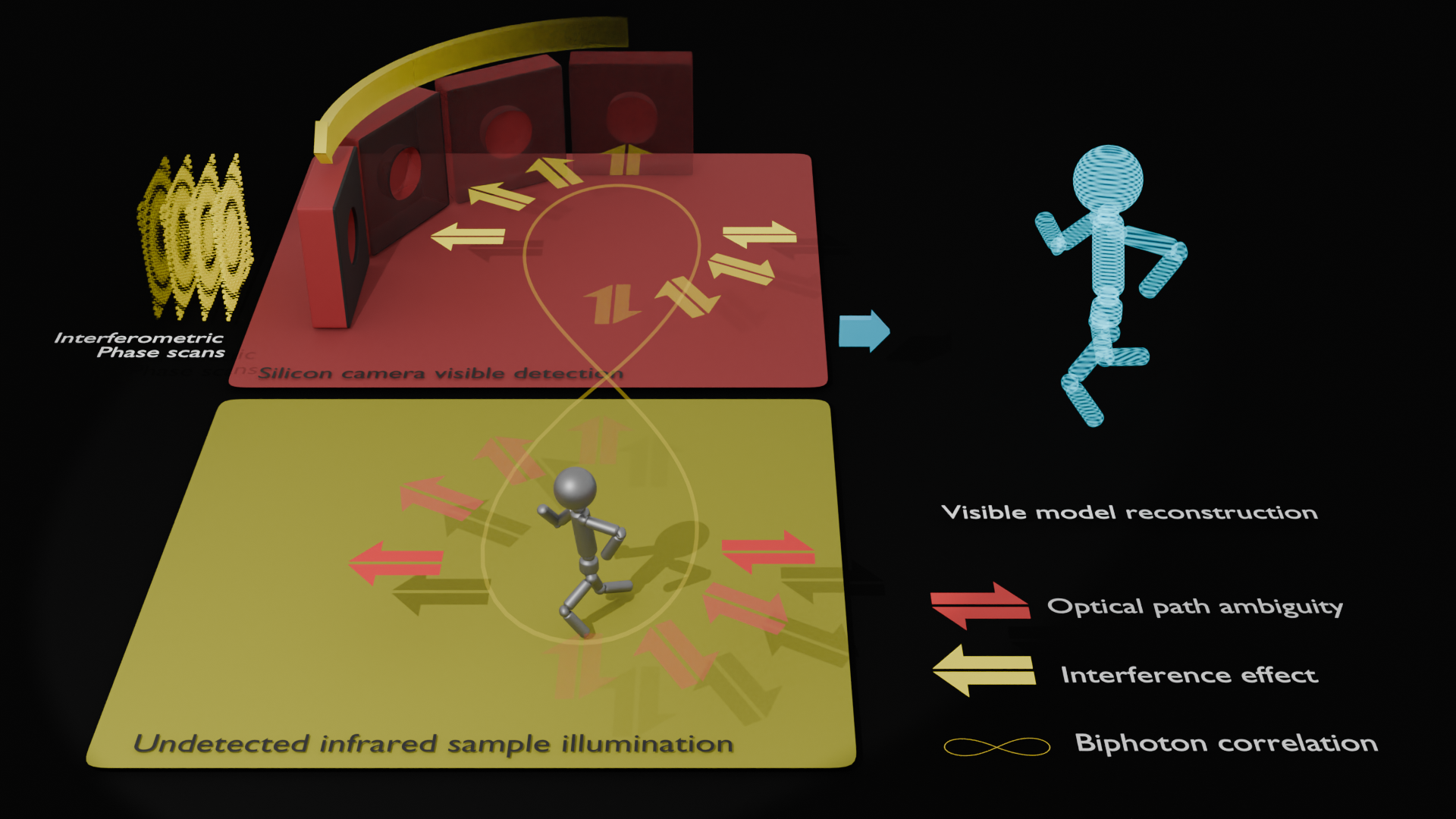}
\caption{\textbf{Quantum undetected optical projection tomography.} An object placed in the infrared arm of the nonlinear interferometer introduces path certainty (bottom left). Which way path information affects interference of the visible beam due to biphoton correlations, thus mapping the object's projection to phase sensitive interferograms on a silicon camera (top left). Compiling projections as a function of object orientation allows the object to be imaged in 3D (right).}
\label{fig:Principle}
\end{figure*}

Mid-infrared light is absorbed by bond-specific vibronic transitions and has been used for many years to analyse chemical mixtures and to produce images with chemical contrast \cite{larkin2017infrared}. Also, infrared light has higher transmission in many materials allowing it to penetrate deeper than visible light \cite{kenry2018recent}. Although a range of infrared OPT scanning techniques have been proposed \cite{marcos2018projection,steinecker2023development,zint2001near}, they are either limited to the near-infrared (wavelengths $\lambda \leq 1.1\,\mu m$) by the detection limits of silicon-based cameras, or, at longer wavelengths, they are seriously limited by the comparatively slow, noisy, expensive and low spatial resolution characteristics of current infrared camera technologies. At wavelengths closer to $\lambda \leq 10\,\mu m$, a key practical challenge is that the signal must be detected against the presence of a large thermal IR background.

Recently the quantum imaging community have actively studied the technique of quantum imaging with undetected photons (QIUP) \cite{lemos2014quantum} that offers a means of circumventing these limitations. It works with pairs of correlated photons, one in the visible and one in the infrared (IR). The quantum nature of the entangled pair means that if an IR photon encounters an object of interest, the interaction is also registered by the visible photon of the pair through a process known as induced coherence \cite{zou1991induced}. This allows the object's information to be transferred from the IR to the visible, where this light can be imaged with higher speed, sensitivity and spatial resolution with inexpensive components. The technique also offers the key advantage of IR thermal noise reduction. The image transfer is via photon correlations so thermal noise is left behind \cite{ma2023eliminating}, and detection noise is then set by visible camera technology. This has the potential to improve the achievable signal-to-noise-ratio (SNR) in infrared imaging by many orders of magnitude. Contrast in the IR (which can contain chemical information) can thus be moved to a wavelength that can be selected for optimal sensitivity, such as the peak quantum efficiency on a silicon camera  \cite{Basset2021,Topfer2022,Kviatkovsky2020,Lindner2020,Arahata2022,Mukai2021,Cardoso2018,Michael2021}. Recently, QIUP was applied to more complex imaging modalities such as optical coherence tomography (OCT) with undetected light \cite{Paterova2018,Rojas-Santana2022,Vanselow2020,Machado2020}.

Here we demonstrate an extension of the QIUP technique: Quantum Undetected Optical Projection Tomography (QUOPT). It combines QIUP and CT concepts to allow sensitive IR 3D imaging. Our system uses room temperature inexpensive off-the-shelf components. We show that, by combining rapid phase scanning \cite{pearce2023practical} with a conventional projection tomography workflow, we can achieve 3D volumetric imaging at wavelengths well beyond the $\lambda \leq 1.1\,\mu m$ cutoff of silicon cameras.

\section*{Experimental Setup}

\begin{figure*}[t]
\includegraphics[width=\textwidth]{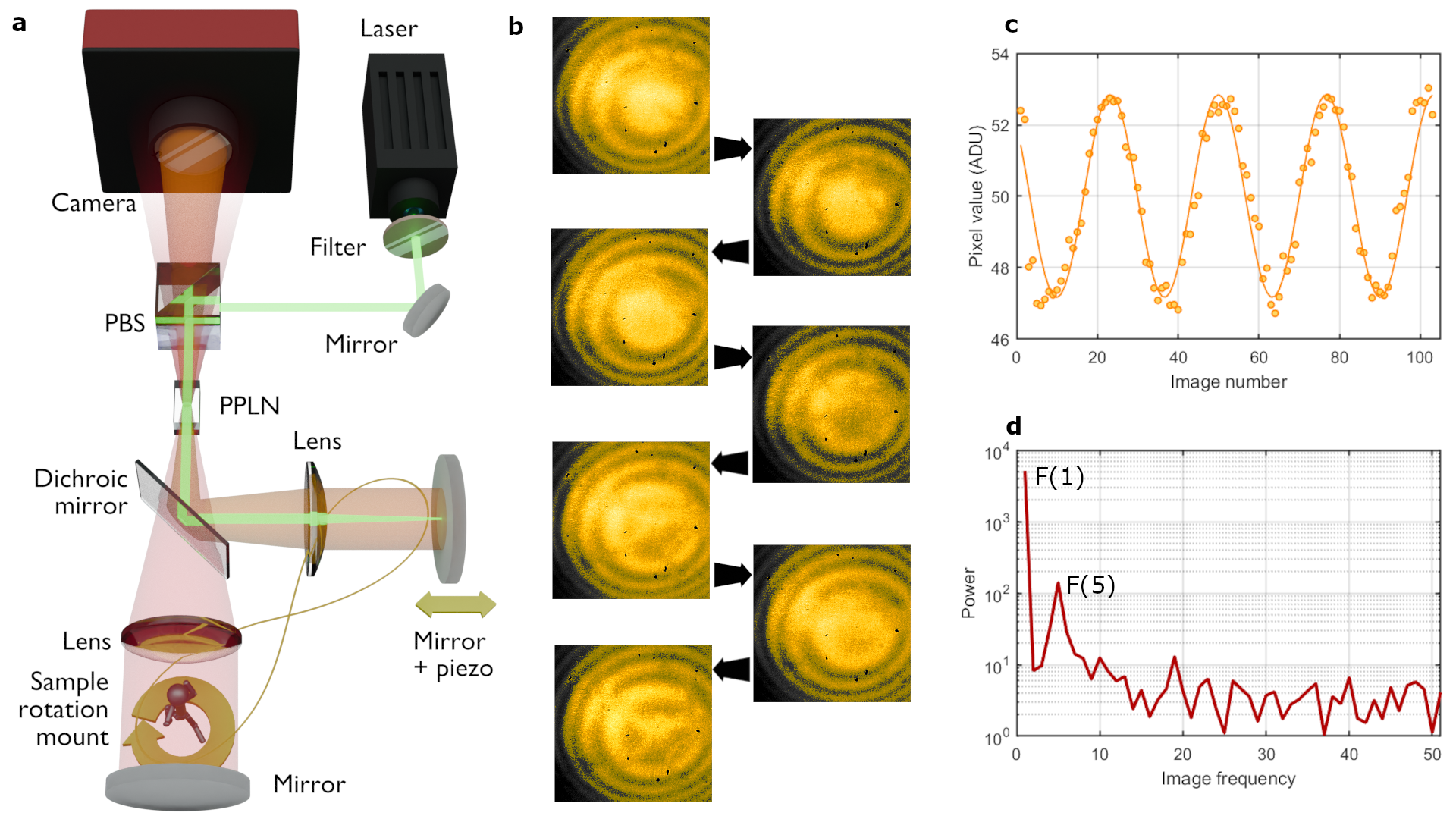}
\caption{\textbf{Quantum undetected optical projection tomography.} \textbf{a} A schematic representation of the quantum imaging with undetected photons (QIUP) setup using momentum correlations. A diode laser pumps a nonlinear periodically-poled lithium niobate (PPLN) crystal to produce entangled pairs of photons, via spontaneous parametric down-conversion (SPDC). A dichroic mirror splits the $\lambda = 1550$\,nm ``idler'' photons from those of the $\lambda = 532\,nm$ pump and $\lambda = 810$\,nm ``signal'', and two mirrors precisely retroreflect all three beams back through the PPLN crystal, rendering photons generated in the first pass of the pump beam indistinguishable from those generated in the second pass. The signal photons are subsequently registered by the silicon camera. The object is placed in the idler beam, on a rotation mount. Its IR transmission image is mapped -- via anti-momentum correlations and the ``induced coherence'' effect -- onto the visibility of the interference fringes that the camera sees in the signal beam as one of the mirrors is scanned with a piezoelectric translation stage. \textbf{b} A sequence of individual `undetected' images taken at different phase delay (mirror) positions; at each pixel the signal varies with changing phase. \textbf{c} An example graph showing how the signal at a single pixel varies with phase delay. \textbf{d} Fourier transform of the data shown in \textbf{c}. The amplitude of the oscillation can be clearly distinguished from the peaks labelled F(1) and F(5) which are used to calculate the fringe visibility (in this example $\sim 5\,\%$).}
\label{fig:schematic}
\end{figure*}

Our QUOPT system is shown schematically in Fig.~\ref{fig:schematic}(a) and described in the Methods section in detail. A green pump beam (532\,nm) passed a nonlinear crystal (PPLN) to generate signal photons at a wavelength that can be detected using a silicon-based complementary metal-oxide-semiconductor (CMOS) camera (810\,nm). The corresponding idler photon wavelength (1550\,nm) lies beyond the detection limit of silicon cameras. Idler photons were retro-reflected back through the crystal via a 4-f imaging system. The signal photons followed the pump through an identical imaging system back through the crystal. The two beam paths were separated via a dichroic mirror. As the pump traversed the crystal a second time there was another possibility of a spontaneous pair generation event, giving rise to interference.

The idler light was then discarded, and the signal light detected on the camera. Interference was recorded on the camera as the mirror that retro-reflects the signal and pump beams was scanned. The resulting ``fringe visibility scan'' (FVS) consisted of a stack of camera images at different fringe phases corresponding to different mirror positions. An example is shown in Fig. 1(b). The interference visibility is defined as 
\begin{equation}
    \mathcal{V} = \frac{N_{max} - N_{min}}{N_{max} + N_{min}}  \, ,
\end{equation}
where $N_{min}$ and $N_{max}$ are the minimum and maximum signals of the interference fringes, which can be calculated for each pixel of the camera to give an image of the visibility across the interfering beam. A typical interference for a single pixel of a FVS is shown in Fig.~\ref{fig:schematic}(c). The visibility $\mathcal{V}$ of the signal is directly linked to the pump, signal and idler transmissions in the interferometer \cite{Gemmell2023} at the corresponding point in the image. While the transmissions of pump and signal beams are fixed, the idler beam transmission is modulated spatially by the object that in turn modulates the signal beam interference visibility at each of the camera pixels. 

An automated rotation mount for the sample enabled FVSs to be completed at multiple illumination angles. The object we used was a figurine made from twisted wire ($\sim$5\,mm in height) which was placed close to the mirror reflecting the 1550\,nm idler beam. A 10\,nm bandpass filter centred at 810\,nm was mounted on the camera to set the corresponding idler wavelength to 1550\,nm. A set of FVSs were taken with the sample rotated a total of 180$^\circ$ in 1$^\circ$ intervals. From each FVS, a visibility image was generated and cropped to ensure a match between the center of the image and the center of rotation. The optical field of view was defined by setting the outer edges of the images to zero. Finally, images were numerically inverted to switch from transmissivity to opacity. 

\begin{figure*}[t]
\includegraphics[width=\textwidth]{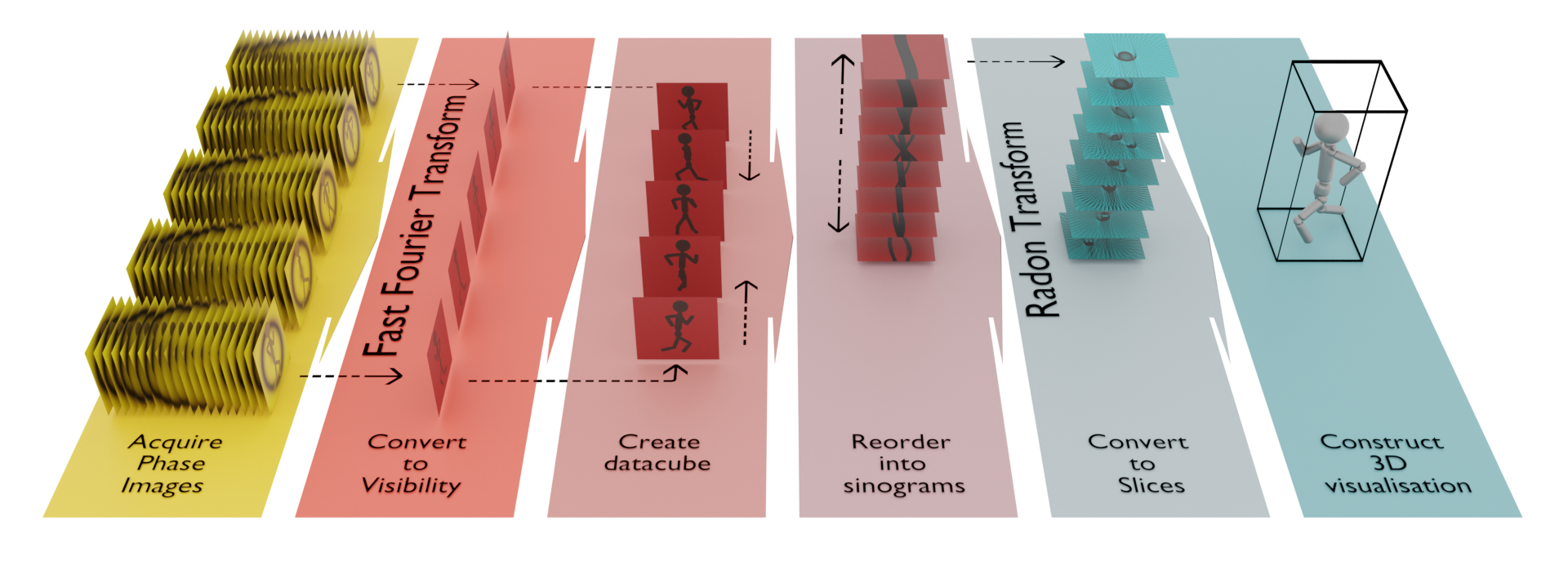}
\caption{\textbf{Image processing workflow.} A representation of the process of using Quantum Imaging with Undetected Photons (QIUP) to generate 3D images via Computed Tomography (CT). A ``Fringe Visibility Scan'' (FVS), consisting of a stack of images at successive phase delays within the interferometer, are each used -- via a fast Fourier transform (FFT) based process -- to generate a transmission image. The process is repeated at multiple rotation angles of the sample to form a datacube. This datacube is reorganised to produce a stack of so-called ``sinograms''. Each sinogram is inverse Radon transformed to produce an image that that maps out the IR absorption across a horizontal slice of the sample. The slices can then be combined in a 3D imaging software to form a visual representation of the volume scanned.}
\label{fig:Process}
\end{figure*}

\section*{\label{sec:level2}Results}

\subsection*{QUOPT Reconstruction Procedure}

Figure 2 illustrates the QUOPT process, from taking raw camera images to reconstructing a 3D volume. Since the frequency of the fringes in the FVS is predefined, their amplitude and visibility can be rapidly computed by using a suitably tailored fast Fourier transform (FFT) \cite{pearce2023practical}. This technique allows the rapid processing and condensation of a large number of phase images into single images displaying the point-wise visibility, amplitude, or phase of the interference. A series of images are acquired from the interferometer with a constant phase delay between each image. To improve the speed of data processing, a one-dimensional FFT performed along the direction of the phase scan can reveal the oscillation frequencies in the scan, allowing isolation of the known frequency of the scanned field. An example FFT is shown in Fig.~\ref{fig:schematic}(d). This operation can be performed for each pixel in the image simultaneously, resulting in a fast and live processing technique to move from stacks of scanned phase images to one image. The visibility can be extracted from the FFTs as
\begin{equation}
    \mathcal{V} \sim 2\Big|\frac{F(n)}{F(1)}\Big|  \, ,
\end{equation}
where $F(n)$ and $F(1)$ are the $n$\textsuperscript{th} and 1\textsuperscript{st} components of the FFT. The $n$\textsuperscript{th} component should correspond to the oscillation frequency, and as such $n$ will depend on the frequency of the scanned field, the scanning step size, and the number of steps. 

This operation can be performed for each pixel simultaneously, generating a transmission image from a FVS image stack effectively in real time. This processing workflow enables the rapid collection of a large number of transmission images that can subsequently be used for volumetric reconstruction by CT.

Many compact QIUP systems are, for simplicity, set up with the sample imaged by light in the far field of the crystal \cite{pearce2023practical,Cardoso2018,Machado2020,Kviatkovsky2020}. In a simplistic picture of the optics, one can imagine that in the scanning field (the idler in this case) a single lens collimates the output of the nonlinear crystal, before transmission through the sample. This sample plane is imaged onto the camera via the signal beams. An FVS then allows an image of the transmission to be extracted. Thus, each row of an image taken with such a system is a direct analogue to CT and OPT where light transmitted through the subject is detected on a camera. A `sinogram' can be built from stacking the same pixel row in a succession of images taken at different sample rotations. Each sinogram (one for each column in the image stack) is then inverse Radon transformed (to convert back to real space). The transformed images are then re-stacked to create a three-dimensional array expressing the transmissivity of the sample at each voxel position. The resultant volumetric slices were stored as separate image files that could be read by 3D rendering software (demonstrated here in Blender \cite{blender_ref}). 

\begin{figure*}[t]
\includegraphics[width=\textwidth]{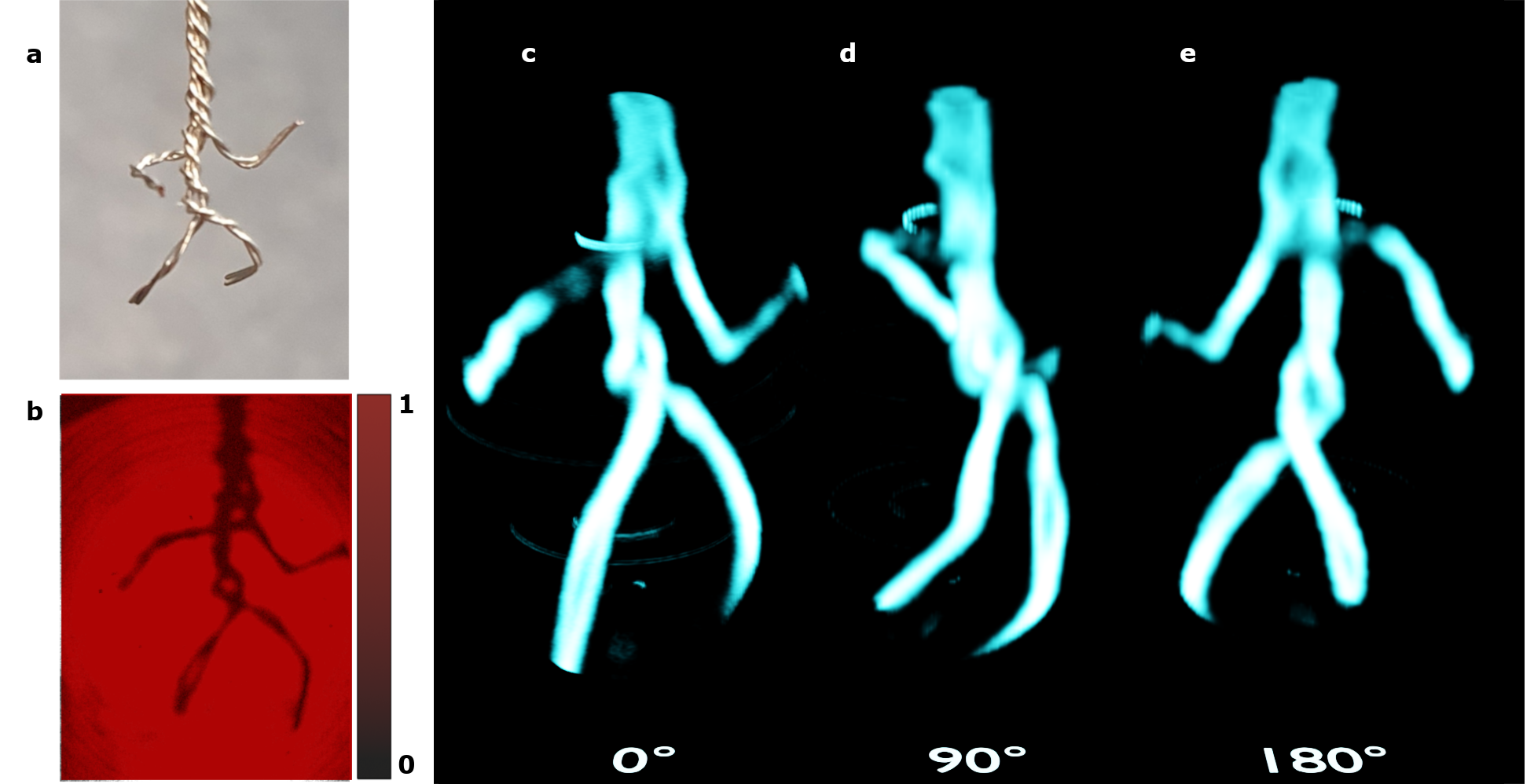}
\caption{\textbf{3D data visualisation.} \textbf{a} A $\sim5$\,mm tall figurine made from twisted wire was scanned to build a 3D model. \textbf{b} An example image of normalised visibility data made from the FFT of a stack of phase images taken at a single rotation angle. \textbf{c-e} Three-dimensional representation of wire figurine visualised and rendered at different angles. }
\label{fig:Result}
\end{figure*}

Figure~\ref{fig:Result}(a) is a photograph of our simple 3D target constructed of twisted wire. An example image of the visibility (and thus transmissivity) of the sample volume at a single target orientation is shown in Fig.~\ref{fig:Result}(b). After taking such images for all rotation angles of the target, we are able to reconstruct the 3D sample, which is rendered by the 3D software. Examples for three different viewing angles are shown in Fig.~\ref{fig:Result}(c-e). A video of the figurine rotating about $360\,^\circ$ can be found in the supplementary material.

\section*{\label{sec:level3}Discussion}

Figure~\ref{fig:Result} shows that it is entirely possible to use the QUOPT technique to measure the transmissivity profile of a 3D volume at difficult to detect wavelengths. This result paves the way for further OPT studies in biological infrared optical windows that may currently be limited by poor and expensive detection technologies. While reliant on the underpinning quantum mechanics, the technique does not require the complex systems typical of quantum optics, thus enabling infrared OPT to become a far more widely used as an investigative tool. 
Additional advancements can be made in the scanning of integrated circuits, where micro-CT is currently used to analyse buried structure\cite{singhal2013micro,suppes2008metrology,lall2014non,rossi2002experimental}. By using wavelengths beyond the band-edge of silicon it will be possible to easily build 3D reconstructions of circuitry without the need for high-power x-ray systems. Extending the wavelength range further into the infrared with alternative nonlinear materials promises the prospect of hyperspectral volumetric infrared spectroscopy, without the need for complex and expensive tuneable infrared laser sources and detectors. Moving the wavelengths further into the infrared ‘molecular fingerprint’ region would allow 3D chemical mapping.

Further work should be done to improve imaging resolution \cite{fuenzalida2022resolution} and field of view to increase the impact on biological imaging fields. The system described here has resolution and field of view limited by its simple design; the 50\,mm focal length lenses coupled with the 10\,mm long nonlinear crystal enable imaging in a cheap and compact design, but at the expense of resolution (for this system estimated to be $\sim300\,\mu$m \cite{pearce2023practical}). Other systems have already shown near diffraction limited imaging capability albeit with more complex optical designs \cite{paterova2020hyperspectral}.

There has been significant work done in the field of OPT to improve the transformation from images of a target taken at different rotations to the reconstruction of a 3D volume \cite{birk2011improved,trull2018comparison,cheddad2011image}, from which this QUOPT technique could benefit.
Further enhancements could be made by making use of recent advances in acquiring QIUP images without the need for phase scanning \cite{pearce2024single,leon2024off}. Such techniques would allow rapid scanning of samples with both phase and amplitude information retrieved which, when combined with the CT scanning shown here, could provide the basis for real-time infrared 3D mapping.

\section*{\label{sec:level5}Methods}
\subsection*{Optical Design}
The experimental setup is based on that presented in \cite{pearce2023practical}. The photon pair source was a 10\,mm long periodically poled lithium niobate crystal with a 0.5 $\times$ 0.5\,mm\textsuperscript{2} aperture. The unpolarised output from a 532nm CW diode laser was prepared spectrally through the use of a bandpass filter (centred at 532\,nm), and in polarisation via reflection from a polarising beamsplitter cube (PBS). The total input power into the crystal after preparation was less than 30\,mW. Signal and idler photons (centred at 810\,nm and 1550\,nm, respectively) were generated by type-0 spontaneous parametric down conversion and separated via a 980\,nm longpass dichroic mirror (the pump beam follows the same path as the signal). Identical 50\,mm (design wavelength 532\,nm) focal length lenses (with different appropriate AR coatings) were placed in the signal and idler paths, one focal length away from the crystal. Two mirrors retroreflected all beams back through the same PPLN crystal. The PBS had a laser line AR coating that acts as an effective dichroic mirror for the pump and signal beams, such that the signal (although vertically polarised) passed with low reflection. A CMOS camera with a longpass filter (cut-off at 700\,nm) recorded the signal beam. A computer-controlled piezo-motor stage attached to the signal/pump retroreflection mirror enabled the phase between the `first crystal' and `second crystal' generation events (and thus the interference fringes) to be scanned. A computer-controlled rotation stage held the sample to be imaged in the path of the idler beam close to the retroreflection mirror. Custom control software enabled the interferometer phase to be stepped, and an image taken at each phase position. The total scan distance was over 4\,$\mu$m. Once the phase scan was complete, the automated program adjusted the rotation stage -- and thus the sample -- by 1\textsuperscript{$\circ$}, and the phase scan was started again. This was repeated for a total of 180\textsuperscript{$\circ$}.

\begin{acknowledgements}
We acknowledge funding from the UK National Quantum Hub for Imaging (QUANTIC, No. EP/T00097X/1), an EPSRC DTP, and the Royal Society (No. UF160475).
\end{acknowledgements}




\bibliography{References}

\end{document}